\documentclass[12pt]{article}
\usepackage{amssymb, amsmath, amsfonts, eurosym, geometry, graphicx, color, setspace, sectsty, comment, footmisc, caption, natbib, pdflscape, array, hyperref, float, verbatim, listings, xcolor, commath, enumitem, xfrac, pgfplots, mathtools, tikz, multirow, diagbox, chngpage, booktabs, tabularx, dirtytalk, indentfirst, comment, setspace, subcaption, rotating, longtable}
\usepackage[normalem]{ulem}
\usepackage{supertabular}

\usepackage[utf8]{inputenc}
\usepackage[affil-it]{authblk}
\spacing{1.1}
\begin{document}

\bibliographystyle{chicago} 
\setcitestyle{round} % set the bracket to be round

%\title{The Social Media Spotlight: An Event Study on ESG Reputational risk %: \\
\title{ESG Reputation Risk Matters: An Event Study Based on Social Media Data
\thanks{This research was funded by Innovate UK project No: 10021201, which funds a consortium led by Kennedys Law LLP and includes Kennedys IQ Limited, University College London, H/Advisors, University of Manchester, and RiskCovered}
}

\author[1]{Maxime L. D. Nicolas\thanks{Corresponding author: m.nicolas@ucl.ac.uk}}
\author[2,3]{Adrien Desroziers}
\author[1,4]{Fabio Caccioli}
\author[1,4]{Tomaso Aste}
\affil[1]{\small{\textit{Department of Computer Science, University College London, UK}}}
\affil[2]{\textit{CES, University of Paris 1 Pantheon-Sorbonne, France}}
\affil[3]{\textit{Department of Management, University of Bologna, Italy}}
\affil[4]{\textit{Systemic Risk Center, London School of Economics and Political Sciences, London, UK}}

\date{}

\maketitle

%%%% Double spacing 
%\doublespacing

\vspace{-1.cm}

\begin{abstract}
\linespread{1}\selectfont

We investigate the response of shareholders to Environmental, Social, and Governance-related reputational risk (ESG-risk), focusing exclusively on the impact of social media. Using a dataset of 114 million tweets about firms listed on the S\&P100 index between 2016 and 2022, we extract conversations discussing ESG matters. In an event study design, we define events as unusual spikes in message posting activity linked to ESG-risk, and we then examine the corresponding changes in the returns of related assets. By focusing on social media, we gain insight into public opinion  and investor sentiment, an aspect not captured through ESG controversies news alone. To the best of our knowledge, our approach is the first to distinctly separate the reputational impact on social media from the physical costs associated with negative ESG controversy news. Our results show that the occurrence of an ESG-risk event leads to a statistically significant average reduction of 0.29\% in abnormal returns.
Furthermore, our study suggests this effect is predominantly driven by Social and Governance categories, along with the \say{Environmental Opportunities} subcategory. Our research highlights the considerable impact of social media on financial markets, particularly in shaping shareholders' perception of ESG reputation. We formulate several policy implications based on our findings.

\noindent

\
%\noindent \textbf{JEL classification}:
\end{abstract}

\thispagestyle{empty}
\clearpage

\newpage

\section{Introduction}

Social media substantially influences corporate perceptions, particularly regarding ESG performance. With the growing interest of investors in ESG factors, it is important to understand the response of shareholders to ESG-related reputation risk (ESG-risk) on platforms like Twitter. However, the existing literature does not provide analyses of this issue. Our study targets this gap, investigating the reaction of shareholders to ESG-risk events through the lens of social media.

The recent development in the ESG landscape has been largely driven by narratives that highlight instances of corporate greed, environmental catastrophes, climate change concerns, and social injustices. Such stories have fuelled the momentum of the ESG movement and triggered demands for increased regulatory action. Consequently, regulators have promoted transparency and completeness in corporate sustainability reporting through policies such as the 2014 Non-Financial Reporting Directive (NFRD) \citep{european2014directive, cicchiello2022non}. This has led to a growing trend towards ESG reporting and disclosure, with more firms communicating their efforts in relation to sustainability through various channels such as sustainability reports, television, and social media.

Public discussions on a firm's ESG matters serves to build a consistent ESG reputation, which can lead to higher financial performance. This process cultivates stakeholder trust and overall brand image to improve market position and investor appeal \citep{benabou2010individual}. In essence, this perspective suggests that well-governed firms can effectively \say{do well by doing good} by implementing transparent communication of explicit ESG practices, while avoiding any contradictory behavior \citep{benabou2010individual, edmans2011does, ferrell2016socially, dyck2019institutional, zhang2022esg}.

A large corpus of research has examined the response of financial markets to the disclosure of negative and positive ESG events \citep{capelle2021shareholders}. This includes studies on accidents \citep{erlac3c6ebib14, erlac3c6ebib58, borenstein1988market}, regulation \citep{erlac3c6ebib47, erlac3c6ebib61, karpoff1993reputational}, and corporate social responsibility related events \citep{erlac3c6ebib42, erlac3c6ebib81, erlac3c6ebib25, farber2009changing}. They find that shareholders sanction negative ESG events and slightly reward positive ones. However, those prior studies focus exclusively on ESG news release and fail to adequately separate the ESG-related reputational effects from the physical costs or profits associated with these events. Indeed, ESG news typically relate to controversies or court decisions involving fees and penalties, which ultimately translate into direct financial costs for the concerned firms. However, it does not reflects public's perception of these companies' ESG performance. To the best of our knowledge, we are the first to fill this gap by examining shareholders' responses to ESG-risk events using an unexplored venue: social media.

Social media has become an increasingly popular way for investors to gather information. It has transformed how accounting information is prepared and disclosed \citep{saxton2012new}, and it offers a form of authenticity, transparency, immediacy, participation, connectedness, and accountability over traditional disclosure media \citep{postman2009socialcorp}. Overall, it provides real-time data and captures a broad range of opinions and perspectives from various stakeholders. In an ESG context, this makes it an efficient tool to capture ESG-related communications and assess ESG reputation. While prior research has explored the use of social media for event studies, there have been no specific focus on examining ESG-risk. An example of related work is \cite{renault2017market}, who used abnormal social media activity to conduct an event study, but the focus was on identifying stock promoters and fraudulent activities.

The present study assesses shareholders' response to ESG-risk events. Specifically, we analyze tweets about firms listed on the S\&P 100 index, collected between 2016 and 2022. From this sample, we extract negative ESG tweets that were not related to specific ESG news or financial announcements, and we define periods of abnormal posting activity about ESG matters as indicators of ESG-risk events.
Subsequently, using an event study, we investigate the associated changes in abnormal returns. Our analysis makes several significant contributions. First, we develop and provide the first extended ESG lexicon based on social media analyses.\footnote{\citet{baier2020environmental}, building on the prior work of \citet{dimson2015active}, proposed an ESG dictionary derived from the textual analysis of 10-K reports from 25 largest firms quoted on the S\&P100. However, these existing dictionaries may not be ideally suited for performing textual analyses of ESG-related content on social media platforms.} Second, we propose a novel method to disentangle ESG-related reputational effects from the physical costs associated with ESG events. Finally, we assess shareholders' response to ESG-risk events, revealing a statistically significant decrease in abnormal returns associated to ESG-risk events. In addition, we find that this effect is more specifically driven by ESG-risks related to Social and Governance categories and events about \say{Environmental Opportunities}. Hence, our results provide evidence of the substantial influence of social media on financial markets, particularly regarding shareholders perception on ESG matters. This study suggests critical policy adjustments, advocating for increased ESG oversight on social media, improved awareness for investors on ESG-risk, and the incorporation of ESG factors in investment strategies to mitigate risk.

\section{Data}

Our research draws from three data sources: financial, social media, and news data. We collected data for the 100 largest U.S. firms listed on the S\&P 100 index for the period from January 2016 to January 2022.\footnote{Although the S\&P 100 technically includes 101 stocks due to dual stock classes for one constituent firm, we omitted the duplicate in our study.} Financial data, i.e. stock price information, is obtained from the daily stock files provided by the Center for Research in Security Prices (CRSP).

We extract social media data via the Twitter Academic Research Application Programming Interface (API) using Python. We use the cashtag\footnote{'Cashtags' are a portmanteau of 'cash' and 'hashtag', composed of a dollar sign and a stock ticker to refer to a particular stock. For instance, \$AAPL is used for Apple.} and the Twitter accounts of each firm to collect relevant tweets. Overall, we have extracted 114 million tweets, corresponding to an average of 1.2 million tweets per firm.

We download ESG controversy news from Refinitiv Eikon, which amounted to 5,390 news headlines for our sample, averaging 54 controversy news items per firm for the period from 2016 to 2022. Additionally, we obtain Earnings Release (ER) dates from Refinitiv's Institutional Brokers' Estimate System (I/B/E/S) dataset, which provided a total of 3,952 dates for the same period.

\section{Methodology}

\subsection{Textual Analysis}

We distinguish ESG and non-ESG related conversation by constructing an ESG lexicon on the three main ESG topics and divided it into 10 categories inspired by MSCI reports, as reported in Table \ref{tab2:lexico}. Consequently, a single tweet can be associated with more than one category or subcategory.

We classify each message's sentiment using the lexicon by \citet{renault2017market}. Then, we construct a daily sentiment score by averaging all the day's message sentiment scores. If the sentiment index is negative (positive) at time $t$, the event is classified as \say{negative} (\say{positive}).\footnote{To address the challenge of identifying negative/positive tweets amid linguistic and contextual subtleties, we apply a 0.05 threshold for classification. This conservative approach categorizes ambiguous or weak sentiment indices as negative. Despite its subjectivity, our analysis maintains robust results across multiple threshold selections around zero.}

Then, we identify an ESG reputation event as a sudden increase in tweet activity at a particular time $t$. We use the ESD identifier \citep{rosner1983percentage} for outlier detection to identify spikes in the volume of messages that contain at least one of the keywords related to the ESG topic. The ESD identifier calculates a score for each data point, based on its deviation from the mean and the standard deviation of the data. Data points with a value higher than a certain threshold are considered outliers.

In particular, an event $x$ at day $t$ is identified if $|x-\bar{X}| \geq z \sigma$. Where $\bar{X}$ is the sample average and $\sigma$ is the sample standard deviation.\footnote{
$\bar{X}$ is the sample average and $\sigma$ is the sample standard deviation.
$$
\bar{X}=\frac{\sum_{i=1}^N x_i}{N}, \sigma=\sqrt{\frac{\sum_{i=1}^N\left(x_i-\bar{X}\right)^2}{N-1}}
$$
where, $N$ is the sample size and $x_{i}$ is the observation $i$ of the sample $N$.
}
$z$ is the sensitivity threshold set to $2$.\footnote{It is common to set $z = 2$ in the ESD identifier because, in normally-distributed data, it is rare to observe a value that deviates more than two standard deviations from the mean (about 4.5\%). Regardless of the probability distribution, when the variance is defined, the Chebyshev's inequality guarantees that such a probability is smaller than 1/4. As a robustness test, we consider $z=3$. The results remained qualitatively similar.} Given the dynamic nature of tweeting behaviors, we account for the non-stationarity of message volume via a 250-day moving window \citep{tumarkin2001news, leung2015impact}. To reduce noise and minor variations, we filter out all events contributing less than 5\% of the total number of messages for a particular firm on a given day, and exclude events with less than 10 tweets \citep{tumarkin2001news, sabherwal2011internet, leung2015impact}.\footnote{The results remained qualitatively similar when we tried different filtering parameters such as a minimum number of tweets of 20.} Furthermore, we adopt a closing-to-closing approach to account for the regular operational hours of stock exchanges,\footnote{We consider all messages sent between 4 p.m. on day $t-1$ and 4 p.m. on day $t$ as relevant to day $t$. This method aligns with the U.S. stock market’s standard trading hours, which include the Nasdaq Stock Market (Nasdaq) and the New York Stock Exchange (NYSE) and run from 9:30 a.m. to 4 p.m., except on stock market holidays.} and enforce a minimum gap of 5 trading days $[-5;5]$ between two events to avoid overlapping.\footnote{If two or more individual outliers fall into our window of interest, they are combined into one single event, where the event is set as being the first outlier in the event window.}

\begin{figure}[H]
\centering
\caption{Number of ESG-reputation events and removed events}
\label{fig:events}
\begin{subfigure}{.45\textwidth}
  \centering
  \caption{Number of events}
  \includegraphics[width=.95\linewidth]{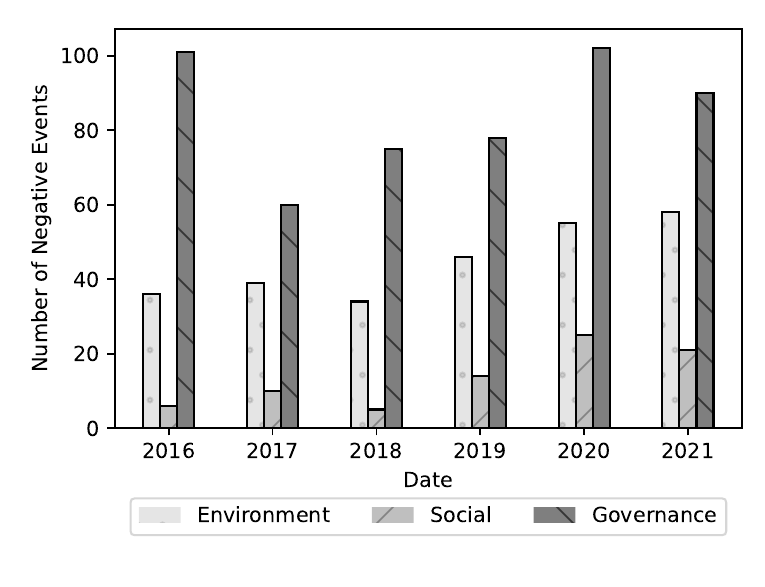}
  \label{fig:sub1}
\end{subfigure}%
\begin{subfigure}{.45\textwidth}
  \centering
  \caption{Number of removed events}
  \includegraphics[width=.95\linewidth]{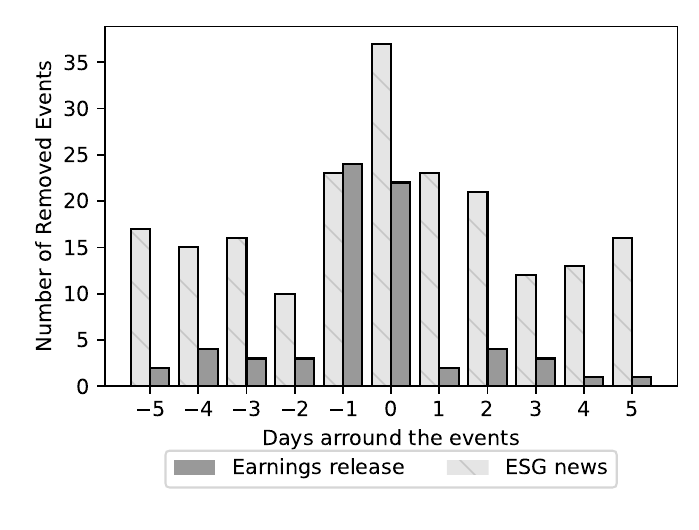}
  \label{fig:sub2}
\end{subfigure}
\end{figure}

To account for the potential endogeneity of social media events to financial markets or firms' activities, we exclude events falling within a $[-5 ; +5]$ days window surrounding earnings releases and ESG controversy news. This process allows us to focus on negative ESG messages unrelated to specific ESG or financial events, capturing the reputational dimension of ESG-risk effectively. Figure~\ref{fig:events} (a) shows the number of events in the sample consisting of  665 ESG-risk events, 253 Environmental-risk, 79 Social-risk, and 495 Governance-risk. Interestingly, we find that the volume of Governance-risk is more substantial. Figure~\ref{fig:events} (b) gives an overview of the distribution of the number of days separating the detected events and the removed event. In this figure, the height of each bar corresponds to the number of events that have been removed, and this is displayed in relation to their distance from news or earnings release dates. The figure shows that a significant amount of financial and ESG news is discussed around ESG-risk events.

\subsection{Event study}

In line with previous research in the economic literature, we employ a market model to estimate the abnormal returns of firms surrounding the disclosure of ESG-risk events. We estimate the abnormal returns as follows:

\vspace{-0.5cm}
$$\operatorname{AR}_{i,t} = \operatorname{R}_{i,t} - \hat{\alpha}_{i} - \hat{\beta} \operatorname{R}_{m,t}$$
\vspace{-1cm}

where $\operatorname{AR}_{i,t}$ represents the abnormal return of firm $i$ at time $t$. $\operatorname{R}_{i,t}$ and $\operatorname{R}_{m,t}$ denote the stock returns and market portfolio returns at time $t$, respectively. The S\&500 index is the benchmark for the market portfolio. Following \citet{ahern2017information} and \citet{perdichizzi2023non}, the estimation window for the market model is set at 120 trading days [-122;-2]. As robustness, we consider an alternative estimation window of 90 trading days to control for potential parameter instability. We determine the standardized average abnormal ($\operatorname{SAARs}$) and standardized cumulative abnormal returns ($\operatorname{SCAARs}$) for a 2-day [-1;0] and 3-day [-1;1] event windows using the formulas:

\hspace{1cm} $\operatorname{SAAR}_{t} =\frac{1}{N} \sum_{i=1}^{N} \operatorname{SAR}_{i,t}$ \hspace{0.5cm} and \hspace{0.5cm} $\operatorname{SCAAR}_{[-n;+n]} =\frac{1}{N} \sum_{i=1}^{N} \sum_{t=x -n}^{x+n} \operatorname{SAR}_{i,t}$

Where, $\operatorname{SAAR}_{t}$ is the standardized average abnormal returns at time $t$, computed as the average of the $\operatorname{SAR}_{i,t}$, i.e. the standardized abnormal return, for each individual firm i on day $t$. $\operatorname{SCAAR}_{[-n;+n]}$ are computed as average sum of daily standardized abnormal returns from day $-n$ to $+n$ around the event day $x$. We follow \citet{boehmer1991event} standardized cross‐sectional test to compute the t-value.

\section{Results}

\begin{figure}[H]
    \centering
    \caption{Standardized Cumulative Average Abnormal Returns around ESG-risk events}\label{fig:figure_caar}
    \label{fig:esg_main}
    \includegraphics[width=0.5\linewidth]{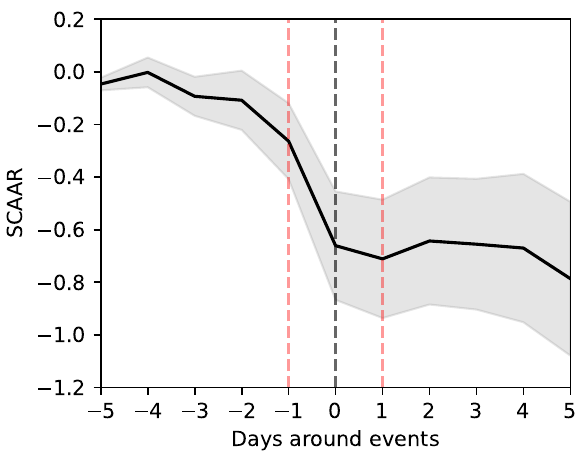} \\
    \footnotesize{The figure show the firms' $\operatorname{SCAAR}$ in the wake of ESG-risk events. \\ The dotted lines represent the event window of interest.}
\end{figure}

\vspace{-0.5cm}
This section assesses shareholders’ response to several types of ESG-risk events. Figure \ref{fig:figure_caar} shows the overall firms' $\operatorname{SCAAR}$ in the wake of ESG-risk events. The null hypothesis asserts that $\operatorname{SCAAR}$ is equal to zero during the event day or period; any significant deviation from random returns implies that events affect a firm's stock prices. Those results indicate a significant negative reaction for ESG-risk events.

\vspace{-0.5cm}

Table \ref{tab:results_esg} examines this effect on an event window of $[-1;+1]$ days which is less subject to capture confounding events. In particular, columns (1) and (2) examine the $\operatorname{SCAAR}$ on an event window of $[-1;0]$ and $[-1;+1]$ days, respectively. Columns (3) to (5) examine the $\operatorname{SAAR}$ from $t-1$ to $t+1$.

The first line presenting the results for the overall ESG-risk events, confirm that shareholders' penalise firms affected by negative ESG communication on social media. Over $[-1;0]$ and $[-1;1]$ days around ESG-risk events, firms stock price decrease on average by 0.29\% to 0.25\%, respectively. This results is also significant on days $t_{-1}$ and $t_{0}$. On the day of the event, ESG-risk events cause a decrease in firm's stock price by 0.29\%. All results are significant at the 1\% threshold, confirming that negative ESG reputation influences shareholders’ perceptions and that ESG-risks are sanctioned.

At the desegregated level, we find that only social- and governance-risk events are penalized by shareholders. Over $[-1;0]$ days around the event, social-risk and governance-risk events reduce firms stock price by -0.36\% and -0.33\% respectively. Results remain qualitatively similar from columns (2) to (4). Conversely, Environmental-risk events do not exhibit significant effects. Those results suggest that shareholders are more sensitive to negative reputation on non-environmental matters.

\begin{table}[H]
    \singlespacing
    \centering
    \caption{Shareholders' response to ESG-risk events}
    \label{tab:results_esg}
    %\resizebox{\textwidth}{!}{%
    \footnotesize{
    \begin{tabular}{lllllll}
\toprule
{} &          $\operatorname{SCAAR}_{[-1;0]}$ &          $\operatorname{SCAAR}_{[-1;1]}$ &           $\operatorname{SAAR}(t_{-1})$ &               $\operatorname{SAAR}(t_0)$ & $\operatorname{SAAR}(t_{+1})$ &     n \\ \midrule
ESG                         &  \textbf{-0.289}$^{***}$ &  \textbf{-0.246}$^{***}$ &  \textbf{-0.117}$^{***}$ &  \textbf{-0.292}$^{***}$ &        -0.018 &   665 \\
                            &                 (-5.265) &                 (-4.628) &                 (-2.749) &                  (-4.540) &      (-0.416) &       \\ \midrule
Environment                 &                    0.001 &                   -0.022 &                    0.067 &                   -0.066 &        -0.038 &   253 \\
                            &                   (0.010) &                 (-0.292) &                   (1.120) &                 (-0.907) &      (-0.578) &       \\
Social                      &  \textbf{-0.355}$^{***}$ &   \textbf{-0.275}$^{**}$ &  \textbf{-0.236}$^{***}$ &   \textbf{-0.266}$^{**}$ &         0.026 &    79 \\
                            &                 (-3.186) &                   (-2.600) &                 (-2.765) &                  (-2.180) &       (0.199) &       \\
Governance                  &  \textbf{-0.325}$^{***}$ &  \textbf{-0.247}$^{***}$ &    \textbf{-0.104}$^{*}$ &  \textbf{-0.356}$^{***}$ &         0.033 &   495 \\
                            &                  (-4.810) &                 (-3.773) &                 (-1.961) &                   (-4.400) &       (0.657) &       \\
\bottomrule
\end{tabular}}%}
    \\
    \resizebox{\textwidth}{!}{%
    \begin{tabular}{lrrrrrrrrrr}
    \multicolumn{9}{p{\textwidth}}{\footnotesize{The table displays the shareholder response following ESG-risk events. The sample includes firms in the S\&P100 index between 2016 and 2022. Column (1) and (2) report the $\operatorname{SCAAR}_{[-1;0]}$ and the $\operatorname{SCAAR}_{[-1;1]}$ which are the standardized cumulative average abnormal returns in percentage from one day before to the ESG-risk events, and from one day before to one day after the ESG-risk events, respectively. Column (3) to (5) report the $\operatorname{SAARs}$ which are the standardized average abnormal returns in percentage from one day before to one day after the ESG-risk events, respectively. Column 6 report the n number of events analyzed. ESG refers to the overall sample of ESG-risk events. Environment, Social and Governance are sub-samples based on ESG-risk events typology. Table \ref{tab2:lexico} provides more details on the typology of ESG-risk events. The table reports results using an estimating window of 120 days. All results include robust standard errors. \citet{boehmer1991event} t-values appear in parentheses. ${ }^{***} p < 0.01, { }^{**} p < 0.05, { }^{*} p < 0.1.$}} \\
    \end{tabular}}
\end{table}
\vspace{-0.5cm}

Delving into our initial findings, Table \ref{tab:results_categories_neg} examines a series of ESG sub-topics. Results suggest that some sub-topics discussed on social media have more effect on financial markets.

Despite the overall lack of impact for the global Environment category, we observe a significant negative effect on firms' value within the \say{Environmental Opportunities}-risk subtopic. Over an event window of $[-1;0]$ days, negative discussions on social media regarding this event class reduce by 0.34\% stock prices. Column (2) and (4), suggest that the effect remains significant on an event window of $[-1;1]$ days and more specifically on the event day $t_{0}$. Furthermore, the absence of effect regarding discussions related to other environmental matters suggests that ESG reputational announcements, i.e. without physical impact, do not alter shareholders perception. Future research should examine if some sectors are not more exposed to some specific ESG-risks.

Regarding Social-risk sub-topics, \say{Product Liability}-risk events appear to lead shareholders negative response. Over a $[-1;0]$ days event window, \say{Product liability}-risk events reduce firms stock prices by 0.32\%. Results remain qualitatively similar and significant at 10\% in Column (2) and (4). \say{Stakeholder Opposition}-risk events exhibit negative abnormal returns over $[-1;0]$ days event windows. The result is significant at 10\%. Specifications regarding $\operatorname{SAARs}$ results are not-significant but consistently negative. Regarding other events type, the lower sample size seem to partially affect the consistency of the results.\footnote{This issue is contingent to the very significant data cleaning that we have conducted using filtering.} \say{Human capital}-risk events are negatively related to abnormal returns but results are only significant at 10\% in column (3). \say{Social Opportunities}-risk only cover 5 events which do not permit to draw meaningful conclusions.\footnote{We choose to present these results as we believe the frequency of these topics on social media is compelling in itself and offers valuable insights for future research.} 

Regarding Governance-risk sub-topics, \say{Corporate Governance}-risk events yields robust results. Over a $[-1;0]$ days event window, stock prices reduce by 0.33\%. From Column (2) to (4), results remain similar and exhibit a stronger market reaction on the event day $t_{0}$. However, \say{Corporate Behavior}-risk do not exhibit a consistent effect.

Table \ref{tab2:results_esg_appendix} presents the robustness test using an estimation window of 90 days to control for potential parameter instability. Results remain qualitatively similar.

\begin{table}[H]
    \singlespacing
    \centering
    \caption{Shareholders' response to ESG-risk events by sub-topics}
    \label{tab:results_categories_neg}
    \resizebox{\textwidth}{!}{%
        \footnotesize{
    \begin{tabular}{lllllll}
\toprule
{} &          $\operatorname{SCAAR}_{[-1;0]}$ &          $\operatorname{SCAAR}_{[-1;1]}$ &           $\operatorname{SAAR}(t_{-1})$ &               $\operatorname{SAAR}(t_0)$ & $\operatorname{SAAR}(t_{+1})$ &     n \\ \midrule
\textbf{Environment}                    &                          &                          &                        &                          &                          &         \\
Climate Change              &                    0.059 &                    0.056 &                    0.021 &                    0.062 &         0.015 &    88 \\
                            &                  (0.633) &                  (0.578) &                  (0.253) &                  (0.615) &       (0.138) &       \\
Natural Capital             &                    0.011 &                   -0.007 &                    0.051 &                   -0.036 &        -0.027 &    79 \\
                            &                  (0.077) &                 (-0.051) &                  (0.404) &                 (-0.292) &      (-0.241) &       \\
Pollution  and  Waste       &                    0.009 &                   -0.012 &                    0.143 &                   -0.129 &        -0.033 &    91 \\
                            &                  (0.083) &                 (-0.113) &                  (1.211) &                 (-1.061) &      (-0.288) &       \\
Environmental Opportunities &   \textbf{-0.337}$^{**}$ &   \textbf{-0.333}$^{**}$ &                   -0.143 &   \textbf{-0.334}$^{**}$ &          -0.100 &    53 \\
                            &                 (-2.521) &                 (-2.157) &                 (-1.202) &                 (-2.177) &      (-0.508) &       \\
\textbf{Social}                    &                          &                          &                        &                          &                          &         \\
Human Capital               &                   -0.231 &                   -0.311 &    \textbf{-0.284}$^{*}$ &                   -0.042 &        -0.213 &    17 \\
                            &                 (-1.135) &                 (-1.574) &                 (-1.799) &                 (-0.174) &      (-1.619) &       \\
Product Liability           &   \textbf{-0.322}$^{**}$ &    \textbf{-0.275}$^{*}$ &                   -0.164 &    \textbf{-0.291}$^{*}$ &        -0.022 &    31 \\
                            &                 (-2.321) &                 (-1.907) &                 (-1.187) &                 (-1.937) &      (-0.098) &       \\
Stakeholder Opposition      &                   -0.282 &     \textbf{-0.380}$^{*}$ &                   -0.154 &                   -0.244 &         -0.260 &    24 \\
                            &                 (-1.676) &                 (-1.954) &                   (-1.200) &                 (-1.371) &      (-0.998) &       \\
Social Opportunities        &     \textbf{0.496}$^{*}$ &                     0.410 &                    0.223 &                    0.479 &         0.009 &     5 \\
                            &                  (2.401) &                  (1.642) &                  (1.834) &                  (1.787) &       (0.024) &       \\
\textbf{Governance}          &                          &                          &                        &                          &                          &         \\
Corporate Governance        &  \textbf{-0.334}$^{***}$ &  \textbf{-0.248}$^{***}$ &   \textbf{-0.109}$^{**}$ &  \textbf{-0.363}$^{***}$ &         0.043 &   465 \\
                            &                 (-4.876) &                 (-3.717) &                 (-1.971) &                 (-4.497) &       (0.839) &       \\
Corporate Behavior          &                   -0.114 &                   -0.122 &    \textbf{-0.326}$^{*}$ &                    0.165 &        -0.051 &    33 \\
                            &                 (-0.594) &                 (-0.665) &                 (-1.702) &                  (0.919) &      (-0.255) &       \\
\bottomrule
\end{tabular}}}
    \resizebox{\textwidth}{!}{%
    \begin{tabular}{lrrrrrrrrrr}
    \multicolumn{9}{p{\textwidth}}{\footnotesize{The table displays the shareholder response following Environmental-, Social- and Governance-risk events sub-topics. The sample includes firms in the S\&P100 index between 2016 and 2022. Column (1) and (2) report the $\operatorname{SCAAR}_{[-1;0]}$ and the $\operatorname{SCAAR}_{[-1;1]}$ which are the standardized cumulative average abnormal returns in percentage from one day before to the ESG-risk events, and from one day before to one day after the ESG-risk events, respectively. Column (3) to (5) report the $\operatorname{SAARs}$ which are the standardized average abnormal returns in percentage from one day before to one day after the ESG-risk events, respectively. Column 6 report the n number of events analyzed. Environment, Social and Governance sub-topics are based on ESG-risk events typology. Table \ref{tab2:lexico} provides more details on the typology of ESG-risk events. Abnormal returns are assessed using an estimating window of 120 days. All results include robust standard errors. \citet{boehmer1991event} t-values appear in parentheses. ${ }^{***} p < 0.01, { }^{**} p < 0.05, { }^{*} p < 0.1.$.}} \\
    \end{tabular}}
\end{table}
\vspace{-1cm}

\section{Conclusion}

To the best of our knowledge, this study is the first to examine how shareholders respond to ESG related reputational risk events and how social media shapes their perception on the matter.
By focusing solely on social media, we distinguish the financial consequences associated with ESG events from the reputational impacts influenced by public opinion on ESG matters.
On the event date of an ESG-risk event, we observe a statistically significant decrease of approximately 0.29\% in abnormal returns.
Furthermore, this effect is stronger for Social and Governance-related risks, specifically \say{Product Liability}, \say{Stakeholder Opposition}, and \say{Corporate Governance}. Environmental-risk events don't have a significant impact on stock prices, unless they are about \say{Environmental Opportunities}.
This suggests shareholders may focus more on the tangible costs of environmental incidents than purely reputational aspects. Hence, our findings underline the powerful role of social media in financial markets, primarily in influencing how shareholders perceive a firm's ESG reputation. 

This study offers several important policy recommendations. Firstly, it highlights the role of both policy makers and firms in strengthening the monitoring of ESG controversies on social media platforms. Secondly, The study also advocates for the integration of ESG factors into investment strategies to mitigate potential ESG-risk exposure. Finally, it underscores the need to amplify investor awareness regarding ESG-risks and opportunities.

Future research could explore sectorial and geographical variations. Indeed, certain sectors may have higher exposure to specific ESG risks, and examining these differences could enhance our understanding of ESG reputation's impact on shareholder perception. Additionally, geographical specificities could offer further insights into this relationship. Lastly, while our current work concentrates on the effects of negative ESG events, it would be valuable to balance this perspective by investigating the potential benefits of positive ESG reputation.

\bibliography{biblio}

\newpage

\section*{Appendix}

\setcounter{table}{0}
\renewcommand{\thetable}{A\arabic{table}}

\begin{table}[H]
    \singlespacing
    \centering
    \caption{Robustness: Shareholders' response to ESG-risk events (estimation window of 90 days)}
    \label{tab2:results_esg_appendix}
    \resizebox{\textwidth}{!}{%
            \footnotesize{
    \begin{tabular}{lllllll}
\toprule
{} &          $\operatorname{SCAAR}_{[-1;0]}$ &          $\operatorname{SCAAR}_{[-1;1]}$ &           $\operatorname{SAAR}(t_{-1})$ &               $\operatorname{SAAR}(t_0)$ & $\operatorname{SAAR}(t_{+1})$ &     n \\ \midrule
ESG                         &  \textbf{-0.286}$^{***}$ &  \textbf{-0.244}$^{***}$ &   \textbf{-0.109}$^{**}$ &  \textbf{-0.296}$^{***}$ &                 -0.018 &   665 \\
                            &                 (-5.047) &                 (-4.395) &                 (-2.549) &                 (-4.468) &               (-0.415) &       \\ \midrule
Environment                 &                    0.005 &                   -0.015 &                    0.072 &                   -0.066 &                 -0.032 &   253 \\
                            &                  (0.062) &                 (-0.197) &                   (1.170) &                 (-0.884) &               (-0.484) &       \\
Social                      &  \textbf{-0.352}$^{***}$ &   \textbf{-0.274}$^{**}$ &  \textbf{-0.236}$^{***}$ &   \textbf{-0.262}$^{**}$ &                  0.023 &    79 \\
                            &                 (-3.228) &                  (-2.58) &                 (-2.849) &                 (-2.156) &                (0.181) &       \\
Governance                  &  \textbf{-0.332}$^{***}$ &  \textbf{-0.255}$^{***}$ &    \textbf{-0.101}$^{*}$ &  \textbf{-0.369}$^{***}$ &                  0.027 &   495 \\
                            &                 (-4.761) &                 (-3.767) &                 (-1.921) &                 (-4.435) &                (0.548) &       \\ \midrule
\textbf{Environment}                    &                          &                          &                        &                          &                          &         \\
Climate Change              &                    0.077 &                    0.083 &                    0.036 &                    0.073 &                  0.036 &    88 \\
                            &                  (0.813) &                  (0.816) &                  (0.419) &                  (0.724) &                (0.329) &       \\
Natural Capital             &                    0.006 &                    0.001 &                    0.053 &                   -0.045 &                 -0.006 &    79 \\
                            &                  (0.039) &                  (0.009) &                  (0.412) &                 (-0.357) &               (-0.052) &       \\
Pollution  and  Waste       &                   -0.010 &                   -0.031 &                    0.143 &                   -0.158 &                  -0.040 &    91 \\
                            &                 (-0.087) &                 (-0.299) &                  (1.189) &                 (-1.285) &               (-0.341) &       \\
Environmental Opportunities &   \textbf{-0.329}$^{**}$ &   \textbf{-0.321}$^{**}$ &                   -0.146 &   \textbf{-0.319}$^{**}$ &                 -0.090 &    53 \\
                            &                 (-2.338) &                 (-2.064) &                 (-1.173) &                 (-2.021) &               (-0.466) &       \\
\textbf{Social}                    &                          &                          &                        &                          &                          &         \\ 
Human Capital               &                   -0.273 &    \textbf{-0.389}$^{*}$ &   \textbf{-0.334}$^{**}$ &                   -0.052 &  \textbf{-0.287}$^{*}$ &    17 \\
                            &                 (-1.331) &                 (-1.967) &                 (-2.367) &                 (-0.202) &                (-2.07) &       \\
Product Liability           &   \textbf{-0.304}$^{**}$ &                  \textbf{-0.252}$^{*}$ &                   -0.131 &    \textbf{-0.299}$^{*}$ &                 -0.007 &    31 \\
                            &                 (-2.241) &                 (-1.697) &                 (-1.005) &                 (-1.928) &               (-0.029) &       \\
Stakeholder Opposition      &    -0.299 &   \textbf{-0.423}$^{**}$ &                   -0.169 &                   -0.254 &                  -0.31 &    24 \\
                            &                 (-1.846) &                 (-2.281) &                 (-1.437) &                 (-1.396) &               (-1.149) &       \\
Social Opportunities        &     \textbf{0.536}$^{*}$ &                    0.457 &                    0.236 &                    0.521 &                  0.034 &     5 \\
                            &                  (2.617) &                  (1.822) &                   (1.930) &                  (1.944) &                (0.094) &       \\
\textbf{Governance}                    &                          &                          &                        &                          &                          &         \\ 
Corporate Governance        &  \textbf{-0.339}$^{***}$ &  \textbf{-0.255}$^{***}$ &    \textbf{-0.103}$^{*}$ &  \textbf{-0.377}$^{***}$ &                  0.039 &   465 \\
                            &                 (-4.765) &                 (-3.653) &                  (-1.880) &                 (-4.495) &                (0.751) &       \\
Corporate Behavior          &                   -0.095 &                   -0.101 &                   -0.302 &                    0.168 &                  -0.040 &    33 \\
                            &                 (-0.513) &                 (-0.549) &                  (-1.63) &                  (0.975) &               (-0.205) &       \\
\bottomrule
\end{tabular}}}
    \\
    \resizebox{\textwidth}{!}{%
    \begin{tabular}{lrrrrrrrrrr}
    \multicolumn{9}{p{\textwidth}}{\footnotesize{The table displays the shareholder response following Environmental-, Social- and Governance-risk events sub-topics. The sample includes firms in the S\&P100 index between 2016 and 2022. Column (1) and (2) report the $\operatorname{SCAAR}_{[-1,0]}$ and the $\operatorname{SCAAR}_{[-1,1]}$ which are the standardized cumulative average abnormal returns in percentage from one day before to the ESG-risk events, and from one day before to one day after the ESG-risk events, respectively. Column (3) to (5) report the $\operatorname{SAARs}$ which are the standardized average abnormal returns in percentage from one day before to one day after the ESG-risk events, respectively. Column 6 report the n number of events analyzed. Environment, Social and Governance sub-topics are based on ESG-risk events typology. Table \ref{tab2:lexico} provides more details on the typology of ESG-risk events. Abnormal returns are assessed using an estimating window of 120 days. All results include robust standard errors. \citet{boehmer1991event} t-values appear in parentheses. ${ }^{***} p < 0.01, { }^{**} p < 0.05, { }^{*} p < 0.1.$.}} \\
    \end{tabular}}
\end{table}

\begin{small}
\renewcommand{\arraystretch}{0.9}
\begin{longtable}[c]{@{}ll@{}}

\caption{Lexicon}\label{tab2:lexico}\\ 

\toprule
\textbf{Topic/Subtopic} &
  \textbf{Keywords} \\* \midrule
\endfirsthead
\multicolumn{2}{c}%
{{\bfseries Table \thetable\ continued}} \\
\toprule
\textbf{Topic/Subtopic} &
  \textbf{Keywords} \\* \midrule
\endhead
\bottomrule
\endfoot
\endlastfoot
\multicolumn{2}{l}{\textbf{ENVIRONMENT}} \\
\multicolumn{2}{l}{\textbf{Climate Change}} \\
Carbon Emissions &
  \begin{tabular}[c]{@{}l@{}}ghg-carbon-methane-nitrous oxide-fluorinated-carbon footprint-\\ decarbonization\end{tabular} \\
  &  \\
\begin{tabular}[c]{@{}l@{}}Climate Change \\ Vulnerability/\\ Resilience\end{tabular} &
  \begin{tabular}[c]{@{}l@{}}climate resilience-climate adaptation-climate mitigation-climate risk-\\ climate vulnerability-climate-smart-climate impact-climate related-\\ climate-sensitive-extreme weather-climate action plan-climate policy-\\ climate strategy-natural disaster-earthquake-hurricane-flood-tornado-\\ tsunamis-wildfire-volcanic eruption\end{tabular} \\
\vspace{-0.2cm} &
   \\
\begin{tabular}[c]{@{}l@{}}Financing\\ Environmental\\ Impact\end{tabular} &
  \begin{tabular}[c]{@{}l@{}}green bonds-esg investing-renewable energy finance-green finance-\\ impact finance-sustainable finance-climate finance-green funds-\\ green loans-carbon credits\end{tabular} \\
\vspace{-0.4cm} &
   \\
\begin{tabular}[c]{@{}l@{}}Product Carbon\\ Footprint\end{tabular} &
  \begin{tabular}[c]{@{}l@{}}product carbon footprint-life cycle-sustainable product-\\ sustainable packaging-low carbon products-eco-friendly products-\\ renewable products\end{tabular} \\
\vspace{-0.4cm} &
   \\
\multicolumn{2}{l}{\textbf{Environmental Opportunities}} \\
\begin{tabular}[c]{@{}l@{}}Opportunities in\\ Clean Tech\end{tabular} &
  \begin{tabular}[c]{@{}l@{}}smart grid-electric vehicle-green technology-smart metering-\\ carbon capture-smart city-sustainable mobility-low carbon transport-\\ fuel cells-advanced materials-air purification-green chemistry-\\ green energy-eco-innovation\end{tabular} \\
\vspace{-0.4cm} &
   \\
\begin{tabular}[c]{@{}l@{}}Opportunities in\\ Green Building\end{tabular} &
  \begin{tabular}[c]{@{}l@{}}sustainable architecture-efficient buildings-leed certification-\\ green building-building durability-building footprint-building energy\end{tabular} \\
\vspace{-0.4cm} &
   \\
\begin{tabular}[c]{@{}l@{}}Opportunities in\\ Renewable Energy\end{tabular} &
  \begin{tabular}[c]{@{}l@{}}renewable energy-wind energy-solar energy-hydropower-\\ biomass energy-geothermal energy-tidal energy-wave energy-\\ energy efficiency-net zero energy-decentralized energy-clean energy-\\ green energy\end{tabular} \\
\vspace{-0.4cm} &
   \\
\multicolumn{2}{l}{\textbf{Natural Capital}} \\
\begin{tabular}[c]{@{}l@{}}Biodiversity \& \\ Land Use\end{tabular} &
  \begin{tabular}[c]{@{}l@{}}biodiversity-land use-ecosystems-habitats-wildlife-endangered species-\\ protected   species-conservation-protected areas-forest management-\\ deforestation-reforestation-land degradation-ecological footprint-\\ ecosystems-wetlands-grasslands-livelihoods-sustainable agriculture-\\ forestry-land conversion\end{tabular} \\
 &
   \\
\begin{tabular}[c]{@{}l@{}}Raw Material\\ Sourcing\end{tabular} &
  \begin{tabular}[c]{@{}l@{}}raw material-supply chain management-sustainable supply chain-\\ ethical sourcing-responsible sourcing-sustainable procurement-\\ eco-friendly sourcing-sustainable resource-sustainable production-\\ resource management-sustainable supply-fair trade-\\ transparency in supply-eco-efficient supply-carbon neutral supply\end{tabular} \\
\vspace{-0.4cm} &
   \\
Water Stress &
  \begin{tabular}[c]{@{}l@{}}water use-water degradation-water protection-clean water-water risk-\\ drought-groundwater-wastewater-desalination-irrigation\end{tabular} \\
\vspace{-0.2cm} &
   \\
\multicolumn{2}{l}{\textbf{Pollution \& Waste}} \\
Electronic Waste &
  electronic waste-electronic recycling \\
\vspace{-0.2cm} &
   \\
\begin{tabular}[c]{@{}l@{}}Packaging\\ Material \& Waste\end{tabular} &
  \begin{tabular}[c]{@{}l@{}}packaging-plastic-waste management-circular economy-waste recycling-\\ upcycling-downcycling-waste-to-energy-waste prevention-zero waste\end{tabular} \\
\vspace{-0.2cm} &
   \\
\begin{tabular}[c]{@{}l@{}}Toxic Emissions \&\\  Waste\\ (Not GHG related)\end{tabular} &
  \begin{tabular}[c]{@{}l@{}}sulfur dioxide-toxic emissions-pollution-chemical emissions-\\ hazardous waste-contaminants-industrial emissions-chemical waste-\\ heavy metals-polychlorinated biphenyls-dioxins-furans-\\ organic compounds-pesticides-inorganic compounds-particulate matter-\\ toxic dust-toxic ash-toxic gases-mercury-cadmium-chromium-arsenic-\\ nickel-selenium-acid rain-ozone depletion-smog-acidification-\\ eutrophication-contamination\end{tabular} \\
\vspace{-0.2cm} &
   \\
\multicolumn{2}{l}{\textbf{SOCIAL}} \\
\multicolumn{2}{l}{\textbf{Human Capital}} \\
\begin{tabular}[c]{@{}l@{}}Human Capital\\ Development\end{tabular} &
  \begin{tabular}[c]{@{}l@{}}human capital-workforce development-talent management-\\ employee training-skills development-talent pool-talent development-\\ leadership development-professional development-employee skills-\\ employee experience\end{tabular} \\
\vspace{-0.4cm} &
   \\
\begin{tabular}[c]{@{}l@{}}Labor\\ Management\end{tabular} &
  \begin{tabular}[c]{@{}l@{}}workforce planning-career development-human resource-\\ talent acquisition-employee retention-workforce diversity-\\ employee satisfaction-employee morale-employee motivation-\\ employee productivity-employee potential-employee performance-\\ employee recognition-employee wellness-employee career-\\ employee resource-employee recognition-employee awards-\\ workforce optimization\end{tabular} \\
\vspace{-0.4cm} &
   \\
\begin{tabular}[c]{@{}l@{}}Supply Chain\\ Labor Standards\end{tabular} &
  \begin{tabular}[c]{@{}l@{}}worker rights-fair labor-decent work-forced labor-child labor-\\ living wage-working conditions-workplace diversity-\\ workplace inclusion-non-discrimination-union rights-\\ sustainable employment-equal opportunity-worker empowerment-\\ harassment\end{tabular} \\
\vspace{-0.4cm} &
   \\
\multicolumn{2}{l}{\textbf{Product Liability}} \\
Chemical Safety &
  \begin{tabular}[c]{@{}l@{}}consumer financial-financial literacy-financial exploitation-\\ financial education-financial counseling-credit protection-\\ financial privacy-financial abuse-predatory lending-investment fraud-\\ insurance fraud-ponzi schemes-money laundering-credit counseling-\\ debt relief-investment advice-credit card debt-loan debt\end{tabular} \\
\vspace{-0.4cm} &
   \\
\begin{tabular}[c]{@{}l@{}}Consumer\\ Financial\\ Protection\end{tabular} &
  \begin{tabular}[c]{@{}l@{}}flood-tornado-tsunamis-wildfire-volcanic eruption-green bonds-\\ esg investing\end{tabular} \\
\vspace{-0.4cm} &
   \\
\begin{tabular}[c]{@{}l@{}}Insuring Health \& \\ Demographic\\ Risk\end{tabular} &
  \begin{tabular}[c]{@{}l@{}}health insurance-demographic risk-life insurance-health risk-\\ medical insurance-disability insurance-long-term care insurance-\\ health cost-health expenses-health spending-health coverage-\\ health access-health equity-health disparities-health claims-\\ health information-health data-health research-health trends-\\ health policies-health reforms-health systems-health administration\end{tabular} \\
\vspace{-0.4cm} &
   \\
\begin{tabular}[c]{@{}l@{}}Privacy \&\\ Data Security\end{tabular} &
  \begin{tabular}[c]{@{}l@{}}data privacy-data security-cybersecurity-information privacy-\\ information security-personal data-sensitive data-confidential data-\\ protected data-data protection-data breaches-data theft-data hacking-\\ data surveillance-data encryption-data management-data compliance-\\ data retention-data backup-data recovery-data storage-data centers-\\ cloud security\end{tabular} \\
\vspace{-0.4cm} &
   \\
\begin{tabular}[c]{@{}l@{}}Product Safety \&\\ Quality\end{tabular} &
  \begin{tabular}[c]{@{}l@{}}product quality-product safety-hazardous materials-product testing-\\ product certification-product standards-product labeling-product recall-\\ product liability-product defect-product failure-product malfunction-\\ product design-product inspection-product durability-product reliability-\\ product performance-product sustainability-product compliance-\\ product traceability-product conformity-product risk-product life cycle\end{tabular} \\
\vspace{-0.4cm} &
   \\
Responsible Invest. &
  \begin{tabular}[c]{@{}l@{}}esg reporting-esg metrics-esg performance-esg integration-\\ esg assessment-esg disclosure-esg management-esg risk assessment-\\ esg sustainability-esg stewardship-esg transparency\end{tabular} \\
\vspace{-0.4cm} &
   \\
\multicolumn{2}{l}{\textbf{Social Opportunities}} \\
Access to Finance &
  \begin{tabular}[c]{@{}l@{}}employee benefits-retirement savings-employee stock ownership-\\ employee stock purchase-employee stock options-\\ employee savings plans-employee pension plans-\\ employee retirement funds-401(k)-defined benefit plans-\\ defined contribution plans-profit sharing-stock bonus plans-\\ employee financial-payroll deduction savings-employee cash balance-\\ employee investment\end{tabular} \\
 &
   \\
\begin{tabular}[c]{@{}l@{}}Access to\\ Health Care\end{tabular} &
  \begin{tabular}[c]{@{}l@{}}health care access-employee health-health benefits-health insurance-\\ health coverage-health services-health support-medical benefits-\\ medical insurance-medical coverage-medical services-health programs-\\ on-site health clinic-social security\end{tabular} \\
\vspace{-0.4cm} &
   \\
\begin{tabular}[c]{@{}l@{}}Opportunities in\\ Nutrition \&\\ Health\end{tabular} &
  \begin{tabular}[c]{@{}l@{}}healthy eating-employee   nutrition-healthy meal-on-site cafeterias-\\ health coaching-health promotion-health and wellness-\\ nutrition awareness-nutrition education-nutritional counseling-\\ nutritional support-wellness program\end{tabular} \\
\vspace{-0.4cm} &
   \\
\multicolumn{2}{l}{\textbf{Stakeholder Opposition}} \\
\begin{tabular}[c]{@{}l@{}}Community\\ Relations\end{tabular} &
  community relations-public relations-stakeholder-local community \\
\vspace{-0.4cm} &
   \\
\begin{tabular}[c]{@{}l@{}}Controversial\\ Sourcing\end{tabular} &
  \begin{tabular}[c]{@{}l@{}}child labor-forced labor-sweatshops-conflict minerals-ethical sourcing-\\ human trafficking-fair trade-human dignity-responsible sourcing-\\ ethical supply chain-human rights-labor exploitation-labor abuses-\\ labor issues\end{tabular} \\
\vspace{-0.4cm} &
   \\
\multicolumn{2}{l}{\textbf{GOVERNANCE}} \\
\multicolumn{2}{l}{\textbf{Corporate Behavior}} \\
Business Ethics &
  \begin{tabular}[c]{@{}l@{}}transparency-fairness-honesty-professionalism-trustworthiness-\\ corporate citizenship-ethic-corporate conscience\end{tabular} \\
\vspace{-0.4cm} &
   \\
Tax transparency &
  corporate tax-fiscal \\
\vspace{-0.4cm} &
   \\
\multicolumn{2}{l}{\textbf{Corporate Governance}} \\
Accounting &
  \begin{tabular}[c]{@{}l@{}}accounting-financial statement-financial reporting-gaap-ifrs-auditing-\\ budgeting-sustainability reporting\end{tabular} \\
\vspace{-0.4cm} &
   \\
Board &
  \begin{tabular}[c]{@{}l@{}}board-corporate governance-director-compensation committee-\\ nomination committee-risk committee-executive-\\ shareholder engagement-chairman-ceo-cfo-coo-independent directors-\\ non-executive directors-executive directors-chairperson-chairwoman-\\ presiding officer-treasurer-financial officer-operating officer\end{tabular} \\
\vspace{-0.4cm} &
   \\
\begin{tabular}[c]{@{}l@{}}Ownership \&\\ Control\end{tabular} &
  \begin{tabular}[c]{@{}l@{}}shareholder rights-shareholder democracy-corporate ownership-\\ minority shareholders-corporate control-ownership concentration-\\ ownership structure-corporate ownership-corporate democracy-\\ corporate elections-corporate vote-corporate representation\end{tabular} \\
\vspace{-0.4cm} &
   \\
Pay &
  \begin{tabular}[c]{@{}l@{}}payment-compensation-incentive-wage-pay benefit-pension plan-\\ stock option\end{tabular} \\* \bottomrule
\end{longtable}
\end{small}

\end{document}